# 3D model-based restoration with positivity constraint using a reduced number of 3D-SIM images

*Cong T. S. Van, Hasti Shabani and Chrysanthe Preza*[*]

Computational Imaging Research Laboratory, Department of Electrical and Computer Engineering,
The University of Memphis, Memphis, TN 38152, USA
*First two co-authors have contributed equally to this work.* [*]cpreza@memphis.edu

**ABSTRACT**

We extend our previous three-dimensional (3D) model-based (MB) approach for 3D structured illumination microscopy (SIM) by introducing a positivity constraint (PC) through the reconstruction of an auxiliary function using a conjugate-gradient method. The performance of the new 3D-MBPC method is investigated with noisy simulation and compared to our previous 3D-MB approach and to the 3D Generalized Wiener filter approach used traditionally for 3D processing of 3D-SIM data. Results show more accurate 3D restoration is possible with the 3D-MBPC method over the other two methods. Moreover, information redundancy in 3D-SIM data is exploited and results obtained with the 3D-MBPC method when the number of raw SIM images is reduced from 15 down to 7 and 5 are promising.

*Index Terms*— Three-dimensional restoration, positivity constraint, structured illumination microscopy, data reduction.

## 1. INTRODUCTION

Three-dimensional structured illumination microscopy (3D-SIM) [1,2], in which the structured illumination (SI) pattern varies laterally and axially, has become one of the most effective optical imaging modalities used in biological investigations because of its optical sectioning and super-resolution capabilities. Since SIM is based on a computational sensing paradigm, computational methods are an integral part of the imaging system modification and have a direct impact on the performance. In our prior work [3], we developed the first 3D model-based (3D-MB) iterative approach for 3D-SIM to restore the final image using a valid forward imaging model that takes into account axial scanning of the sample. This method was shown to provide more accurate results in noisy simulation than the non-iterative 3D generalized Wiener filter (3D-GWF), the standard deconvolution approach proposed by Gustafsson [1], at the expense of longer computation time. In general, model-based iterative approaches allow reconstruction of information outside the spatial frequency bandwidth, which is set by the optical system [4]. Moreover, these approaches provide robustness to noise through regularization, flexibility in applying a positivity constraint thereby avoiding unrealistic negative values, joint estimation of the SI patterns and the corresponding parameters and data-acquisition reduction by taking advantage of redundancies in the forward images. In this paper, we develop, based on the 3D-MB framework, a new method that enforces positivity of the solution and investigate its performance. In many cases introducing the positivity *a priori* information improves the performance of the inverse method, as it was shown in speckle-based SIM [5]. The joint Richardson-Lucy algorithm recently applied to SIM was shown to produce a restored image that is positive provided that the initial guess does not have any negative values [6]. Other model-based algorithms used in SIM based on least squares optimization included a positivity constraint through either a regularization term [7] or through the reconstruction of an auxiliary function [5,8,9] as we do here.

Redundancy of information in raw SIM images has been investigated recently to speed up data acquisition thereby reducing phototoxic effects and bleaching. The minimum number of images required for successful reconstruction in 2D-SIM, in which the SI pattern does not change axially, was first introduced by Heintzmann [10] from the perspective of information theory. Later, Orieux et al. [11] and Dong et al. [12] provided a realization of the use of 4 images instead of the traditional 9 raw SIM images in 2D processing of 2D-SIM data. In Orieux et al. [11], the 4 images include the wide-field (WF) image and 3 raw SIM images from 3 different orientations. However, Dong et al. [12] proposed complementary phases instead of three phases in 2D-SIM and using 4 raw SIM images: 2 from one orientation with 0 and $\pi$ phase (which result in the WF image when summed) and 2 from two other orientations. Reducing the number of images in 2D-SIM down to 3 raw SIM images (one from each of 3 different orientations) was proposed by Strohl et al [13] as an underdetermined optimization problem. In the case of three-wave interference 3D-SIM (3W-SIM) data, the same reduction, i.e. using 4 images (the WF image and 3 raw SIM images) instead of the traditional 15 raw SIM images [1] has also been applied, however only for 2D processing of a single axial section [14]. Here, we show results obtained with our new 3D model-based with positivity constraint (3D-MBPC) method using 7 and 5 out of 15 raw 3D-SIM images, in which 3D processing of the entire volume is performed.

## 2. MODEL AND METHOD

In a 3D-SIM system, as described in [3], the intensity in the 3D image recorded using axial scanning can be modeled as:

$$g(\boldsymbol{x}, z) = (Ao)(\boldsymbol{x}, z) := \sum_{k=1}^{K}[o(\boldsymbol{x},z)j_k(\boldsymbol{x})] \otimes [h(\boldsymbol{x},z)i_k(z)], \quad (1)$$

where $\boldsymbol{x} = (x, y)$ and $z$ are the transverse and axial coordinates respectively; $o(\boldsymbol{x},z)$ is the density distribution of fluorophores within the sample; $i_k(z)$ and $j_k(\boldsymbol{x})$ are the axial and lateral functions of the SI pattern, respectively; $h(\boldsymbol{x},y)$ is the point spread function of the imaging system and $(Ao)(\boldsymbol{x},z)$ is the convolution operator defined in Eq. (1).

One can use the 3D model-based (3D-MB) approach to restore the original 3D image of the sample $o(\boldsymbol{x},z)$ as described in [3]. Since the desired fluorescence distribution of the underlying sample is non-negative, we suggest, in this paper, a 3D model-based with positivity constraint (3D-MBPC) approach using the conjugate-gradient descent algorithm to reconstruct $o(\boldsymbol{x},z)$ through the auxiliary function:

$$o(\boldsymbol{x},z) = \zeta^2(\boldsymbol{x},z). \quad (2)$$

Starting from the cost function with respect to $o(\boldsymbol{x},z)$,

$$F(o(\boldsymbol{x},z)) = \sum_{l=1}^{L} \|g_l^{mes}(\boldsymbol{x},z) - g_l(\boldsymbol{x},z)\|^2, \quad (3)$$

where $\|.\|$ is the $l^2$-norm and $g_l^{mes}$, $g_l$ are the $l^{th}$ 3D recorded image and model prediction, respectively, the gradient with respect to $\zeta(\boldsymbol{x},z)$ is computed as:

$$\nabla_\zeta F = \nabla_o F \frac{\partial o}{\partial \zeta}$$
$$= -4\zeta(\boldsymbol{x},z) \sum_{l=1}^{L}\left(A^\dagger(g_l^{mes} - g_l)\right)(\boldsymbol{x},z), \quad (4)$$

where $A^\dagger$ is the adjoint operator of $A$, which is the cross-correlation operator and $L$ is the total number of SIM images. Similar to [3], the conjugate-gradient descent algorithm is applied using the recursive relation:

$$\hat{\zeta}_n = \hat{\zeta}_{n-1} + \alpha_n d_n, \quad (5)$$

where the updating direction $d_n$ is:

$$d_n = \nabla_n + \gamma_n d_{n-1}, \text{ and } \gamma_n = \frac{\langle \nabla_n | (\nabla_n - \nabla_{n-1}) \rangle}{\|\nabla_{n-1}\|^2}, \quad (6)$$

and the step size $\alpha_n$ is determined at each iteration $n$ by minimizing the cost function with respect to $\alpha_n$, i.e., $F(\zeta + \alpha_n d)$.

## 3. DATA REDUCTION

In prior works of model-based SIM restoration using a reduced number of raw SIM images (by taking into account information redundancy in the raw images), a WF image of the same field of view was used to replace some of the SIM images, as it provides low frequency information and prevents ambiguity [11,12,14]. The WF image is available either from direct observation by changing the imaging modality from SIM to WF [11, 14] or by summing SIM images recorded with selected complimentary phases along one orientation of the SI pattern [12]. Here, we first propose using 7 out of the 15 images traditionally used in 3W-SIM by selecting 5 images from one orientation (to ensure we have the same information as the WF image) and 2 images from two other orientations. We also reduce the number of images further from 7 to 5 (by having 3 images from one orientation and 2 from two other orientations), based on the conjugate symmetry property in the frequency domain [11].

## 4. RESULTS

To investigate the performance of the 3D-MBPC approach with and without data reduction and to compare it to the performance of the 3D-MB approach, we use the same synthetic object as in [3]. The synthetic object is simulated on a 512×512×512 grid of a cube with a side equal to 6.4 $\mu m$. This object has an outer spherical shell and inner spherical beads (Fig. 1e). The diameter and thickness of the spherical shell are 3 $\mu m$ and 200 nm, respectively, while the diameter of each bead is 150 nm and the closest distance between two neighboring beads is 175 nm. We apply the forward imaging model Eq. (1) to the object to obtain the raw SIM data and then down-sample it to a 256×256×256 grid simulating the effect of the CCD camera. In addition, Poisson noise [15] is applied to the simulated data at the SNR level of 15 dB. To compare the results of different approaches, the restored 3D images are normalized using the $l^2$-normalization and negative values in the 3D-MB and 3D-GWF results are then set to 0. The mean square error (MSE) and the structural similarity index measure (SSIM) [16] are used to compare the performance of the approaches quantitatively.

The simulation is for the 3W-SIM system [1] using a 63X/1.4 NA oil-immersion (refractive index n = 1.515) lens with an excitation wavelength $\lambda$ = 515 nm. Three orientation angles ($\theta$ = 0°, 60°, 120°) of the SI pattern are used to obtain isotropic resolution, with 5 SI pattern phases $\varphi$ shifted by a $2\pi/5$ step starting with $\varphi = 0$ rads. The following parameters are chosen: $u_m = 0.8 u_c$, $\frac{w_m}{u_m} = \frac{w_c}{u_c} = \frac{NA^2}{2n\lambda} = 0.231$, where $u_m$ and $w_m$ are the lateral and axial modulating frequencies, respectively, while $u_c$ and $w_c$ are the lateral and axial cut-off frequencies, respectively. Note that for this simulated system, the lateral resolution limit of the conventional WF microscope is $d = 0.61\lambda/\text{NA}$ = 224 nm, which is larger than the smallest distance between two neighboring beads in the object, while on the other hand the predicted SIM lateral resolution limit, $d_{\text{SIM}} = d/(1+u_m/u_c) = $ 125 nm, is smaller than the smallest object distance.

In the first simulation study, we compare the result from 3D-MBPC to the ones from the standard 3D-GWF [1] and the

3D-MB [3] approaches (Fig. 1). The Wiener parameter in the 3D-GWF method is chosen to be 0.01. For both the 3D-MBPC and 3D-MB methods, the initial guess is set equal to the WF image, which is obtained from the sum of the 5 raw images from one orientation ($\theta = 0°$) and the number of iterations is 150. Fig. 1 shows the results from the three approaches. The $xy$-section images (top row of Fig.1) indicate that both the 3D-MB and 3D-MBPC results have better lateral distinction between two neighboring beads than the 3D-GWF (as evidenced clearly in the intensity line plots), but the 3D-MBPC result approximates the intensity better than the 3D-MB result because there is no energy lost in the solution from the negative restored values as in the 3D-MB result. The $xz$-section images (middle row of Fig. 1) show that a better optical sectioning capability is achieved by the 3D-MBPC when considering the axial diameter and the intensity of the restored inner beads (see axial intensity line plots). Moreover, the MSE and SSIM metrics computed between the restored and true object intensities (see table in Fig. 1) show that the 3D-MBPC result provides a more accurate restoration than the 3D-MB and 3D-GWF results.

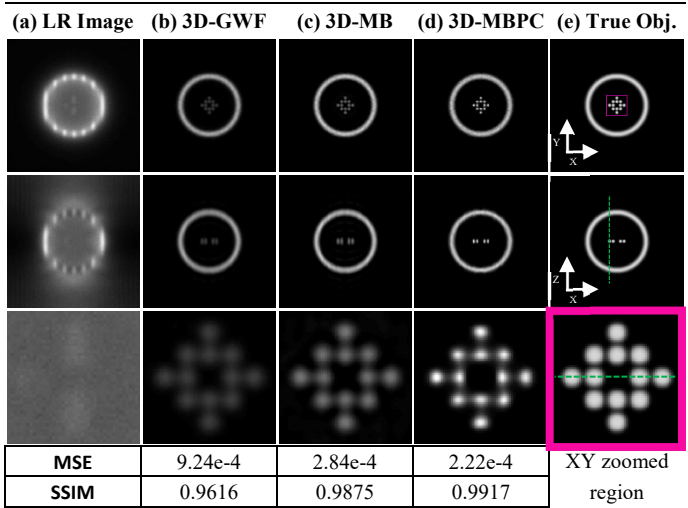 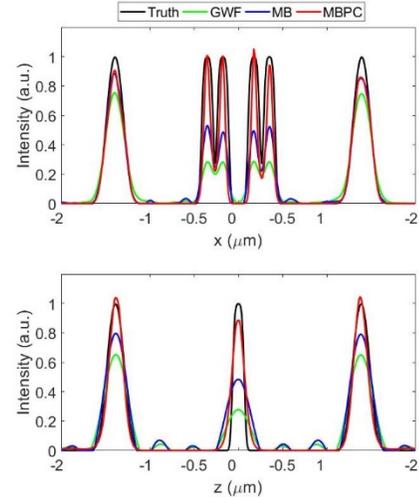

**Fig. 1.** Comparison of 3D restorations for 3D-SIM. (a) One of the 15 low resolution (LR) raw 3W-SIM images (on a 256-cubic grid, zoomed-in) simulated using a 63x1.4NA oil lens at a 515nm wavelength and SNR = 15 dB. Restoration on a 512-cubic grid: (b) 3D-GWF, with Wiener parameter 0.01; (c) 3D-MB at 150 iterations; and (d) 3D-MBPC at 150 iterations; (e) True object. The right-hand-side panels show lateral (top) and axial (bottom) intensity profiles taken along the green dashed lines marked in the bottom and middle rows of (e), respectively. Images are displayed on the same scale [0, 1.2] and the negative values are removed from both the 3D-GWF and 3D-MB results.

In the second simulation study, the data reduction (described in Section 3) is exploited using both the 3D-MBPC and 3D-MB methods and Fig. 2 compares results obtained with and without data reduction. The chosen 7 images contain 5 raw images from one orientation $\theta = 0°$ and 2 raw images at $\varphi = 2\pi/5$ from two other orientations. The chosen 5 images contain 3 raw images from one orientation $\theta = 0°$ at $\varphi = 0, 2\pi/5, 4\pi/5$ and 2 raw images at $\varphi = 2\pi/5$ from two other orientations. When we choose 5 out of 15 raw images, the initial guess is set equal to the sum of the 3 raw images from one orientation ($\theta = 0°$), which is different than the WF image. From the $xy$-section images and the metrics table, the results obtained with data reduction are seen to be very similar to the results without data reduction, for both the 3D-MB and 3D-MBPC methods. As evident from the results, some artifacts are visible in the zoomed-in images when only 5 out 15 images are used, nevertheless the beads are still resolvable (Fig. 3c and 3f). As expected, there is a tradeoff between data acquisition and restoration accuracy when data reduction is applied. While data acquisition time decreases linearly with a decreasing number of images, we have found that more iterations are needed when data reduction is used.

## 5. CONCLUSION

The 3D-MBPC approach developed here for 3D-SIM introduces positivity *a priori* information about the desired fluorescence intensities of the underlying object through the restoration of an auxiliary function using a conjugate-gradient method. As expected, the positivity constraint improves the solution to the 3D inverse problem in 3D-SIM. Results from noisy simulation obtained with the 3D-MBPC method, with and without data reduction, are shown to provide more accurate restoration than the ones obtained with the 3D-MB and the 3D-GWF methods. Compared to the 3D-MB and 3D-GWF methods (Fig. 1), the 3D-MBPC method achieves 0.4% and 3.1% restoration improvement in terms of SSIM, and 21.8% and 76% improvement in terms of MSE, respectively, when all 15 raw images are used. With the reduction of data by using 7 or 5 out of 15 raw images, compared to the 3D-MB method, the 3D-MBPC method achieves 1.6% restoration improvement in terms of the SSIM, and a 25.9% improvement in terms of the MSE when 7 images are used (the MSE values are the same when 5 images are used). For the effect of data reduction on the 3D-MBPC method, with 7 out of 15 raw images, the MSE increases by 19.8% and the SSIM decreases by 0.16%, while with 5 out of

15 raw images, the MSE increases by 99.5% and the SSIM decreases by 0.82%. We note that the SSIM metric measures the difference in the shapes of the results while the MSE metric measures the difference in the intensities of the results. Therefore, the % difference in the SSIM values reported for the three approaches is not as much as the corresponding % difference in the MSE values. Data reduction in 3D-SIM is desirable and although previous studies showed results with 2D processing of single section images, here we have shown proof-of-concept results for 3D processing of volumes acquired with 3W-SIM without the need to acquire a widefield image. These results suggest that further investigation is warranted.

|  | (a) MB - 15 | (b) MB – 7 | (c) MB - 5 | (d) MBPC-15 | (e) MBPC-7 | (f) MBPC-5 |
|---|---|---|---|---|---|---|
| XY | 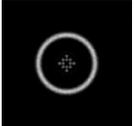 | 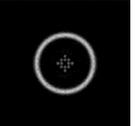 | 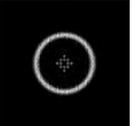 | 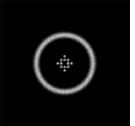 | 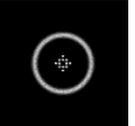 | 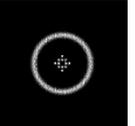 |
| XZ | 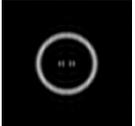 | 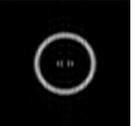 | 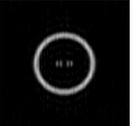 | 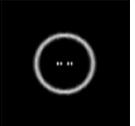 | 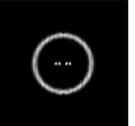 | 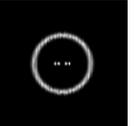 |
| XY zoomed region | 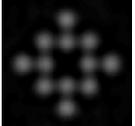 | 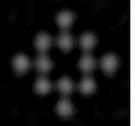 | 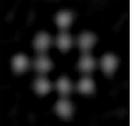 | 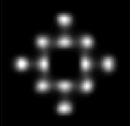 | 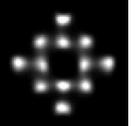 | 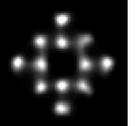 |
| MSE | 2.84e-4 | 3.59e-4 | 4.43e-4 | 2.22e-4 | 2.66e-4 | 4.43e-4 |
| SSIM | 0.9875 | 0.9743 | 0.9689 | 0.9917 | 0.9901 | 0.9836 |

**Fig. 2.** Comparison of 3D-MB and 3D-MBPC restoration using data reduction with a different number of raw 3W-SIM images (on a 256-cubic grid) simulated using a 63x1.4NA oil lens at a 515 nm wavelength and SNR=15 dB. Restoration on a 512-cubic grid: (a) 3D-MB with all 15 raw images after 150 iterations; (b) 3D-MB with 7 out of 15 raw images after 200 iterations; (c) 3D-MB with 5 out of 15 raw images after 200 iterations; (d) 3D-MBPC with all 15 raw images after 150 iterations; (e) 3D-MBPC with 7 out of 15 raw images after 200 iterations; (f) 3D-MBPC with 5 out of 15 raw images after 200 iterations. Images are displayed on the same scale [0, 1.2] and the negative values are removed from the 3D-MB result. 7 and 5 out of 15 images are selected as described in Section 4.

## 6. ACKNOWLEDGEMENT

This work is supported by the National Science Foundation (NSF, DBI 1353904; PI: CP) and Herff fellowship to HS from the University of Memphis (UoM). We thank our collaborators G. Saavedra (University of Valencia, Spain) and A. Doblas (UoM) for technical discussions.